\documentclass[aps,preprint,showpacs]{revtex4}

\usepackage{amssymb}
\usepackage{amsmath}
\usepackage{epsfig}
\usepackage{epstopdf}
\usepackage{bm}
\usepackage{graphicx,epsfig}
\usepackage{mathrsfs}
\usepackage{dcolumn}
\usepackage{color}
\usepackage{natbib}
\usepackage{CJK}
\def\be{\begin{equation}}
\def\ee{\end{equation}}
\def\bea{\begin{eqnarray}}
\def\eea{\end{eqnarray}}

\allowdisplaybreaks[2]

\begin{document}

\title{Dynamic localization of light from a time-dependent effective gauge field for photons}
\author{Luqi Yuan and Shanhui Fan}
\affiliation{Department of Electrical Engineering, and Ginzton
Laboratory, Stanford University, Stanford, CA 94305, USA}

\date{\today }

\begin{abstract}
We introduce a method to achieve three-dimensional dynamic
localization of light. We consider a dynamically-modulated
resonator lattice that has been previously shown to exhibit an
effective gauge potential for photons. When such an effective
gauge potential varies sinusoidally in time, dynamic localization
of light can be achieved. Moreover, while previous works on such
effective gauge potential for photons were carried out in the
regime where the rotating wave approximation is valid, the effect
of dynamic localization persists even when the counter-rotating
term is taken into count.
\end{abstract}

\pacs{42.60.Da, 63.20.Pw, 41.20.Jb}

\maketitle

The effect of dynamic localization is of fundamental importance in
understanding coherent dynamics of a charged particle in a
periodic potential. When such a charged particle is in addition
subjected to a time-harmonic external electric field, the
wavefunction of the particle can become completely localized
\cite{dunlap86,holthaus92}. This effect has been studied in a
number of systems
\cite{platero04,nazareno97,dignam02,wang04,madureira04,eckardt05,creffield10,tsuji11,singh12,longhi12,bulgakov14},
and has been demonstrated in experiments involving Bose-Einstein
condensate or optical lattices \cite{eckardt09,madison98}.

Localization of photon, especially in full three dimensions, is of
great practical and fundamental importance for the control of
light \cite{john87,garanovich12}. Hence dynamic localization of
photon is of significance as well. Photon is a neutral particle,
thus there is no naturally occurring time-harmonic electric field
that couples to photon. To achieve dynamic localization of photon,
one therefore needs to synthesize an effective electric field. Up
to now, extensive experimental and theoretical works has focused
on light propagation in a waveguide array, where the effect of
dynamic localization manifests  by analogy as the cancellation of
diffraction when the array is modulated in space along the
propagation direction
\cite{longhi05,longhi06,szameit09,szameit10,joushaghani12,crespi13}.
There has not been however any demonstration of a true
three-dimensional localization of light in a photonic structure
undergoing time-dependent modulation.

In this letter, we show that the concept of photonic gauge
potential provides a mechanism to achieve dynamic localization of
light in full three dimensions. It has been theoretically proposed
\cite{fang12} and experimentally demostrated
\cite{fangprb13,tzuang14,li14} that when the refractive index of a
photonic structure is modulated in time sinusoidally, the phase
the modulation corresponds to an effective gauge potential for
photon states \cite{fang12np,fang13,fang13oe,lin14}. Ref.
\cite{fang12,fangprb13,tzuang14,li14,fang12np,fang13,fang13oe,lin14}
utilized this correspondence to create a spatially inhomogeneous,
but time-invariant gauge potential distribution, in order to study
effects associated with an effective magnetic field for photons,
including the photonic Aharnov-Bohm effect
\cite{fang12,fangprb13,li14}, and the photonic analogue of the
integer quantum hall effect \cite{fang12np}. In contrast, here we
create a gauge potential that is spatially homogenous or periodic,
but temporally varying. We show that such a time-dependent gauge
potential naturally leads to a time-varying effective electric
field for photons, which can be used to create three-dimensional
dynamic localization of light. Moreover, while Ref.
\cite{fang12,fang12np,fang13,fang13oe,lin14} have only considered
the regime where the rotating wave approximation is valid, here we
show that such dynamic localization persist even when the
counter-rotating term is taken into account.

\begin{figure}[b]
\includegraphics[width=10cm]{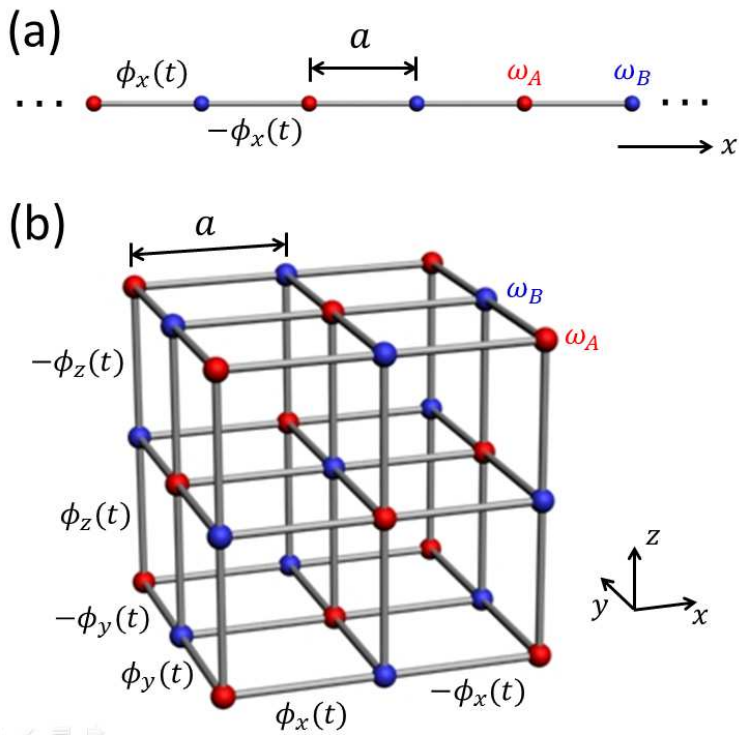}
\caption{(color online) A one-dimensional (a) and a
three-dimensional (b) photonic resonator lattice where two kinds
of resonators with frequency $\omega_A$ (red dots) and $\omega_B$
(blue dots). The nearest-neighbor coupling is dynamically
modulated and the phase of the coupling constant modulation itself
can be time-dependent with the signs being flipped between two
neighboring bonds. The lattice is assumed infinite in all
directions.} \label{lattice}
\end{figure}

We start with the same model system as discussed in details in
Refs. \cite{fang12np,fang13}, consisting of either a
one-dimensional or three-dimensional photonic resonator lattice as
shown in Fig. \ref{lattice}. The lattice consists of two types of
resonators ($A$ and $B$) with frequencies $\omega_A$ and
$\omega_B$ respectively. The Hamiltonian of the system is
\begin{equation}
H = \omega_A \sum_n a^\dagger_n a_n + \omega_B \sum_n b^\dagger_n
b_n  + \sum_{\langle mn \rangle}V\cos(\Omega t + \phi_{mn}(t))
(a^\dagger_m b_n+b^\dagger_n a_m), \label{ham1}
\end{equation}
where $V\cos(\Omega t +\phi_{mn}(t))$ is the coupling strength
between the nearest neighbor resonators. $\Omega = \omega_A -
\omega_B$. $\phi_{mn}$ is the phase of the couplng strength
modulation. In this paper, we will consider the situations where
such modulation phase itself is modulated in time, and refer to
such modulation of the phase $\phi_{mn}$ as the \textit{phase
modulation}. $a^\dagger$ ($a$) and $b^\dagger$ ($b$) are the
creation (annihilation) operators in the $A$ and $B$ sublattice,
respectively.

In the limit $V \ll \Omega$, the rotating wave approximation (RWA)
is valid. Therefore, we can simplify the Hamiltonian and rewrite
it in the rotating frame \cite{fang12np}
\begin{equation}
H = \sum_{\langle mn \rangle} \frac{V}{2} \left(
e^{-i\phi_{mn}(t)} c^\dagger_m c_n + e^{i\phi_{mn}(t)} c^\dagger_n
c_m \right), \label{hamrwac}
\end{equation}
where $c_{m(n)} = e^{i\omega_{A(B)}t}a_m(b_n)$. In general such a
system has a dynamic effective gauge field \cite{fang12np}
\begin{equation}
\vec A^{\mathrm{eff}}_{mn} = \hat l_{mn} \phi_{mn}(t) / a,
\label{gaugeA}
\end{equation}
where $\hat l_{mn}$ is a unit vector and $a$ is the distance
between two near-neighbor sites. Here however, we choose the
modulation phases such that in Eq. (\ref{hamrwac}), all bonds
along the same direction have the same phase. e.g. all bonds along
the $x$-direction has the same phase $\phi_x(t)$. In the
three-dimensional case, $\phi_y(t)$ and $\phi_z(t)$ are similarly
defined. Since the phases are uniform in space, the system has
zero effective magnetic field.

\begin{figure}[t]
\includegraphics[width=10cm]{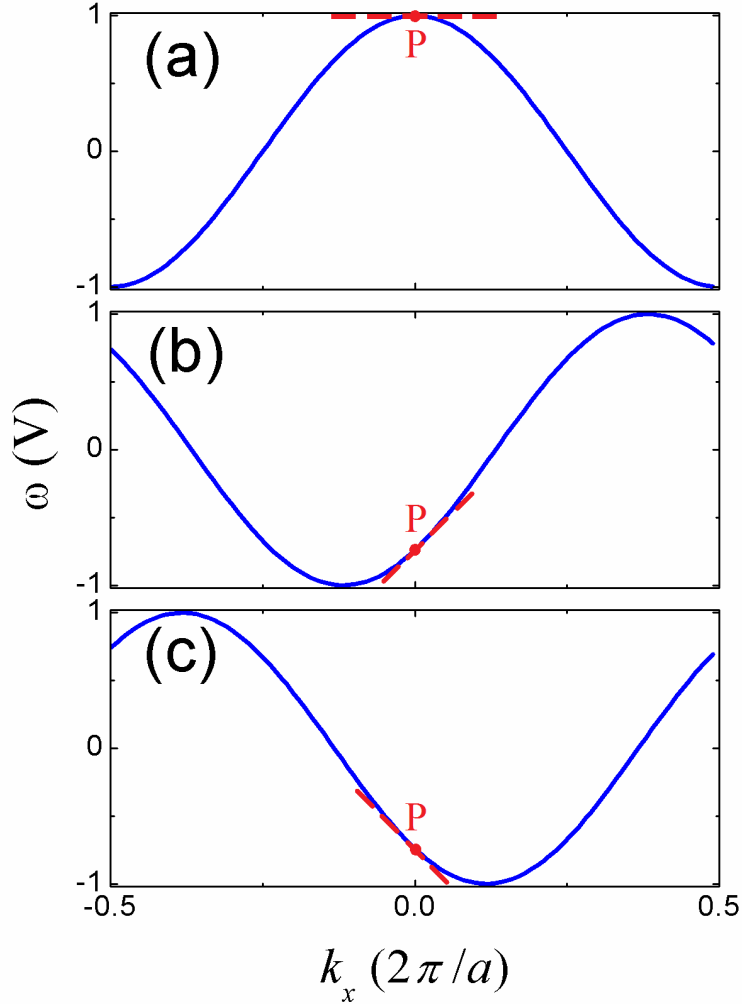}
\caption{(color online) The band diagram for the one-dimensional
lattice as shown in Figure \ref{lattice}(a), with (a) $\phi_x =
0$, (b) $\phi_x=2.40483$, and (c) $\phi_x=-2.40483$. Red dashed
lines indicate the slope at point P.}\label{figband}
\end{figure}

We now show that with a proper choice of the time-dependency of
these phases, we can achieve dynamic localization. As an
illustration we consider the one-dimensional case in some details.
The three-dimensional case naturally follows. In the
one-dimensional case, as an intuitive analysis, we can write the
Hamiltonian (\ref{hamrwac}) in the wavevector space
($\bm{k}$-space)
\begin{equation}
H = \sum_{k_x} V c^\dagger_{k_x} c_{k_x} \cos[k_x a - \phi_x(t)] .
\label{hamrwack}
\end{equation}
Hence the system has an instantaneous photonic band structure
$\omega (k_x) = V \cos[k_x a - \phi_x(t)] = V \cos[(k_x-A_x)a]$.
The instantaneous bandstructures at three different values of
$\phi_x$ are shown in Figure \ref{figband}. The effect of a
spatially uniform photonic gauge potential is a shift of the
bandstructure in $\bm{k}$-space (\cite{fang13,lin14}). Since the
structure maintains translational invariance, the wavevector $k_x$
is a conserved quantity throughout the modulation process. The
group velocity of the wave packet of the photon with wavevector
$k_x$ is given by
\begin{equation}
v_g (k_x) = \frac{\partial \omega (k_x)}{\partial k_x} = - V a
\sin[k_x a - \phi_x(t)] . \label{vgrwa}
\end{equation}
At different values of $\phi_x$, the group velocity at the same
wavevector can have either positive or negative signs. (Figure
\ref{figband}).

To demonstrate dynamic localization, we choose a phase modulation
of the form $\phi_{x}(t) = \alpha_{x}\cos(\omega_M t)$, where
$\alpha_{x}$ and $\omega_M$ are the amplitude and the frequency of
the phase modulation, respectively. Thus, the average group
velocity over one phase modulation period $2\pi/\omega_M$ is
\begin{equation}
\langle v_g (k_x) \rangle= - V a \sin(k_x a) \mathrm{J_0}
(\alpha_x) , \label{vgaverwa}
\end{equation}
where $\mathrm{J_0}$ is the zeroth-order Bessel function. by
choosing $\alpha_x$ be to a zero of $\mathrm{J_0}$, (we denote one
of such zero as $\alpha$ below), the average group velocity is
zero for all $k_x$. Thus all wavepackets of the system become
localized, signifying the presence of dynamic localization.
Importantly, the condition for dynamic localization here is
related to the strength of the phase modulation, and independent
of the phase modulation frequency $\omega_M$.

\begin{figure}[t]
\includegraphics[width=10cm]{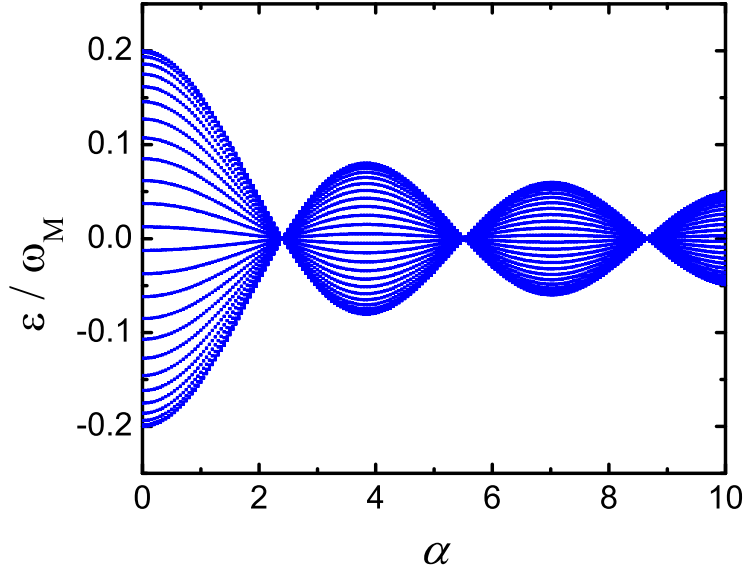}
\caption{(color online) The quasienergies as a function of
$\alpha$, for the Hamiltonian of Eq. (\ref{hamrwack}), with
$\phi_x(t) = \alpha \cos (\omega_M t)$. Here we choose $V = 0.2
\omega_M$. Each curve corresponds to a different wavevector $k_x$,
in the range $-\pi/a < k_x < \pi/a$. }\label{figBCrwa}
\end{figure}

We confirm the intuitive analysis above based on the instantaneous
bandstructure, by a rigorous numerical calculation of the Floquet
eigenstates of Hamiltoian in Eq. (\ref{ham1}). In this numerical
analysis, we use the Hamiltonian of Eq. (\ref{hamrwack}), and
directly compute the quasi-energy $\varepsilon$ at each $k_x$,
following the same procedure as in Ref.
\cite{holthaus92,sherley65,samba73}. The resulting $\varepsilon$
as a function of phase modulation strength $\alpha$, for different
$k_x$'s, are plotted in Figure \ref{figBCrwa}. At each $\alpha$,
the range of the values of the quasi-energy indicates the
bandwidth of the quasi-energy bandstructure.  The onset of the
dynamic localization corresponds to the collapse of the bandwidth.
In Figure \ref{figBCrwa}, we indeed observe the collapse of
bandwidth when the phase modulation strength approaches each of
the zero's of $\mathrm{J_0}$.

For the study of electronic dynamic localization, the effect of a
time-varying electric field is typically described through the use
of a spatially non-uniform \emph{scalar} potential, as described
by a Hamiltonian \cite{dunlap86,holthaus92}
\begin{equation}
\tilde H = \sum_{\langle mn \rangle} \frac{V}{2} \left(
c^\dagger_m c_n + c^\dagger_n c_m \right) - \sum_n
n\alpha\omega_M\sin(\omega_M t) c^\dagger_n c_n,
\label{hamrwaceff}
\end{equation}
In contrast, we have used a \emph{vector} potential that is
spatially periodic. Our Hamitonian of Eq. (\ref{hamrwac}) is in
fact equivalent to (Eq. (\ref{hamrwaceff})) by a gauge
transformation:
\begin{equation}
|\Psi\rangle = \sum_n v_n c^\dagger_n|0\rangle \rightarrow
|\tilde\Psi\rangle = \sum_n \tilde v_n c^\dagger_n|0\rangle =
\sum_n v_n e^{i\theta_n} c^\dagger_n|0\rangle ,
\label{eqgaugetrans}
\end{equation}
where $|\Psi\rangle$ satisfies the Schr\"{o}dinger equation
$i\frac{\partial}{\partial t}|\Psi\rangle=H|\Psi\rangle$ or $i\dot
v_n= \frac{V}{2}\left[ e^{-i\alpha\cos(\omega_M t)} v_{n+1} +
e^{i\alpha\cos(\omega_M t)} v_{n-1}\right]$. With a gauge choice
of $\theta_n = -n\alpha\cos(\omega_M t)$, the gauge-transformed
state $|\tilde\Psi\rangle$ then satisfies
\begin{equation}
\begin{aligned}
i\frac{\partial}{\partial t}|\tilde\Psi\rangle & =\sum_n i \dot
v_n e^{i\theta_n} c^\dagger_n|0\rangle - \sum_n v_n \dot \theta_n
e^{i\theta_n} c^\dagger_n|0\rangle \\
& = \frac{V}{2} \sum_n \left[ e^{-i\alpha\cos(\omega_M t)} v_{n+1}
+ e^{i\alpha\cos(\omega_M t)} v_{n-1}\right] e^{i\theta_n}
c^\dagger_n|0\rangle - \sum_n v_n \dot \theta_n e^{i\theta_n}
c^\dagger_n|0\rangle \\
& = \frac{V}{2} \sum_n \left[ v_{n+1}e^{i\theta_{n+1}} +
v_{n-1}e^{i\theta_{n-1}}\right] c^\dagger_n|0\rangle - \sum_n
n\alpha\omega_M \sin(\omega_m t) v_n e^{i\theta_n} c^\dagger_n|0\rangle \\
& = \tilde H |\tilde\Psi\rangle,
\end{aligned}
\end{equation}
where $\tilde H$ is given in Eq. (\ref{hamrwaceff}). Therefore,
the two Hamiltonians of Eqs. (\ref{hamrwac}) and
(\ref{hamrwaceff}) are indeed equivalent to each other, as they
are related by a gauge transformation. Similar gauge
transformation has been used in the study of waveguide array
\cite{longhi05}. Certainly, a time-varying gauge potential for an
electron is related to an electric field applied on the electron.
Here we have shown that a time-varying effective gauge potential
for a photon also analogously produces an effective electric field
applied on the photon.

Unlike the waveguide array approach, where the effect of photonic
dynamic localization manifests through an analogy as the
cancellation of diffraction in a static structure, in our approach
here one can directly achieve dynamic photon localization in all
three dimensions. We consider the Hamiltonian of Eq. (\ref{ham1})
for the three-dimensional lattice as shown in Figure
\ref{lattice}(b). We choose the phase modulation $\phi_{x,y,z}(t)
= \alpha\cos(\omega_M t) $.  The intuitive derivation of dynamic
localization condition (Eqs. (\ref{vgrwa})-(\ref{vgaverwa})) can
then be straightforwardly generalized to full three-dimension.
Full three-dimensional dynamic localization is achieved provided
that the modulation strength above is chosen to be a zero of the
$\mathrm{J_0}$, for all choices of phase modulation frequency
$\omega_M$.

\begin{figure}[t]
\includegraphics[width=12cm]{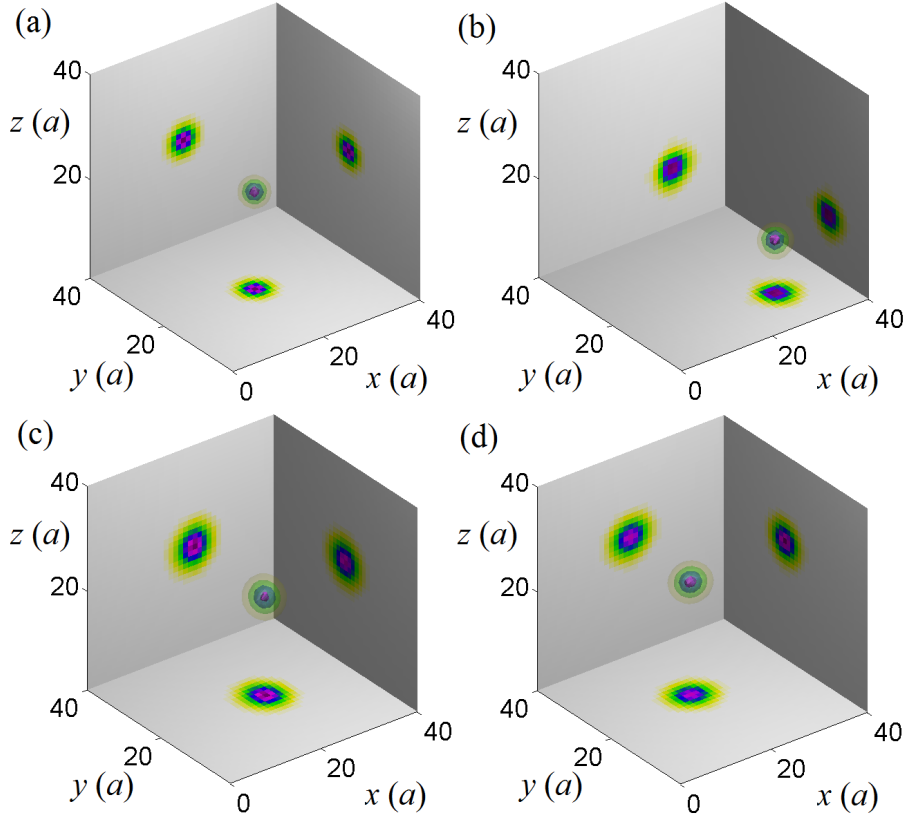}
\caption{(color online) Propagation of a photon wavepacket in a
$40a\times40a\times40a$ three dimensional lattice. (a) The initial
condition at $t=0$, with $x_0 = y_0 = z_0 = 20$ $a$ and $k_x =
-k_y = -k_z = - 1.283$ $a^{-1}$; (b) The wave packet at $t=1.25$
$a/c$ with no phase modulation; (c) and (d) The wave packet at
$t=1.25$ $a/c$ and $t=5$ $a/c$, respectively, with phase
modulation. The parameters of the phase modulation are
$\alpha=2.40483$ and $\omega_M = 1.5$ $c/a$. The coupling strength
between the resonators is $V = 2.4 \pi$ $c/a$. }\label{fig3D}
\end{figure}

Similar to the one-dimensional case, the intuitive derivations for
dynamic localization for three-dimension can be confirmed by a
rigorous Floquet analysis showing band collapse. Instead, here we
provide the evidence of full three-dimensional dynamic
localization, by a direct simulation of photon dynamics in a
$40a\times40a\times40a$ three dimensional lattice. The simulation
is done by solving the coupled-mode equation \cite{fang13oe}
\begin{equation}
i d |\Psi(t)\rangle/dt = H(t) |\Psi(t)\rangle. \label{motion}
\end{equation}
Here $|\Psi\rangle = \left[ \sum_m v_m(t)a^\dagger_m + \sum_n
v_n(t)b^\dagger_n \right] |0\rangle$ gives the photon state with
the amplitude at site $m(n)$ described by $v_{m(n)}(t)$.  $H(t)$
is the time-dependent Hamiltonian of Eq. (\ref{ham1}). The initial
wave packet of the photon at $t=0$ has the form $|\Psi(0)\rangle =
\prod_{\eta=x,y,z} \exp[-(\eta-\eta_0)^2/w^2 + ik_\eta \eta]$,
where $(x_0,y_0,z_0)$ is the center of wave packet with waist $w$.
The results are plotted in Figure \ref{fig3D}. In the absence of
phase modulation, Figure \ref{fig3D}(a) shows the initial wave
packet of the photon. The wave packet propagates freely in the
space with time and reaches to the corner of the lattice at $t =
1.25$ $a/c$ (see Figure \ref{fig3D}(b)). In contrast, in the
presence of phase modulation with a choice of the amplitude
$\alpha=2.40483$ and frequency $\omega_M = 1.5$ $c/a$, the wave
packet of the photon is localized near its initial position
throughout the entire duration of the simulation. This is
demonstrated in Figs. \ref{fig3D}(c) and (d), which show the wave
packet's positions at $t=1.25$ $a/c$ and $t=5$ $a/c$,
respectively. The simulation here provides a direct visualization
of the dynamic localization process in three dimensions.

Up to this point we have used the rotating wave approximation for
the Hamitonian in Eq. (\ref{ham1}). Previous discussions on the
photonic gauge field in this Hamiltonian have all assumed the
rotating wave approximation. On the other hand, in the
experimental demonstration of the photonic gauge field one often
uses electro-optic modulation of refractive index \cite{tzuang14}.
In many electro-optic modulations, the strength of the modulation,
as measured in $\delta n/n \times \omega_0$, where $n$ is the
refractive index of the structure, $\delta n$ is the index change,
and $\omega_0$ is the operating frequency, can be much larger than
the modulation frequency $\Omega$ on the order of a few GHz,
therefore it is of importance to understand the validity of gauge
potential concept beyond the rotating wave approximation. Here we
show that the dynamic localization effect persists even in the
regime where the rotating wave approximation is not valid.

We provide the results in one-dimension. The generalization to
three-dimension is straightforward. For the treatment beyond the
rotating wave approximation, we again starts by providing an
intuitive treatment based on the instantaneous band-structure, we
then confirm the intuitive treatment through an exact numerical
analysis of the Floquet bandstructure. The Hamiltonian
(\ref{ham1}) can be written in $\bm{k}$-space as:
\begin{equation}
\begin{aligned}
& H = \sum_{k_x} \left(\omega_A a^\dagger_{k_x}a_{k_x} + \omega_B
b^\dagger_{k_x}b_{k_x} \right) + \sum_{k_x} V
a^\dagger_{k_x} b_{k_x} \times \\
&   \left(e^{i\Omega t} \cos[k_x a +\phi_x(t)] + e^{-i\Omega t}
\cos[k_x a -\phi_x(t)]\right)+ h.c.. \label{hamnonrwak}
\end{aligned}
\end{equation}
Perform the transformation $c_{k_x} =
e^{i\omega_{A(B)}t}a_{k_x}(b_{k_x})$, we obtain
\begin{equation}
H = \sum_{k_x} V c^\dagger_{k_x} c_{k_x} \left\{ \cos[k_x a -
\phi_x(t)] + \cos(2\Omega t)\cos[k_x a + \phi_x(t)] \right\} .
\label{hamnonrwack}
\end{equation}
We notice that the first term is the same as Eq. (\ref{hamrwack})
and the second term is the counter-rotating term. From Eq.
(\ref{hamnonrwack}) we can straightforwardly obtain the
instantenous bandstructure and hence the instanteuous group
velocity at a wavevector $k_x$, since the Hamiltonian in the
presence of counter rotating term is still periodic in real space.
Again, assuming that the modulation phase $\phi_x =
\alpha\cos(\omega_M t)$. The average group velocity over one phase
modulation period ($2\pi/\omega_M$) is
\begin{equation}
\langle v_g (k_x) \rangle  = - V a \sin(k_x a) \mathrm{J_0}
(\alpha) - Va \frac{\omega_M}{2\pi}\int^{2\pi/\omega_M}_0 dt
\cos(2\Omega t)\sin[k_x a-\alpha\cos(\omega_M t)].
\label{vgavenonrwa}
\end{equation}
To facilitate analytic calculation, we assume that
\begin{equation}
2\Omega = n\omega_M, \label{eqcondition}
\end{equation}
where $n$ is a positive integer, the second term in Eq.
(\ref{vgavenonrwa}), denoted as $\langle v_g (k_x)
\rangle_{\mathrm{CR}}$, can be calculated analytically as
\begin{equation}
\langle v_g (k_x) \rangle_{\mathrm{CR}} = Va  \times \left\{
{\begin{array}{*{20}c}
   (-1)^{m+1} \sin(k_x a) \mathrm{J_n} (\alpha) & n=2m \\
   (-1)^{m} \cos(k_x a) \mathrm{J_n} (\alpha) & n=2m+1 \\
\end{array}} \right. .\label{vgaveCR}
\end{equation}
By choosing $\alpha=2.40483$, which corresponds to $\mathrm{J_0}
(\alpha)=0$, the first term in Eq. (\ref{vgavenonrwa}) vanishes.
And the correction due to the second term can be made arbitrarily
small by choosing a sufficiently large $n$ in Eq.
(\ref{eqcondition}), i.e. by choosing the phase modulation
frequency to be sufficiently small as compared to the frequency of
coupling strength modulation. Thus, dynamic localization can still
be accomplished in the regime where rotating wave approximation no
longer applies. This result can be straightforwardly generalized
to there-dimension. Three-dimensional dynamic localization should
occur, when $2 \Omega = n \omega_M$, provided that all bonds along
each direction has the same phase
$\phi_{x,y,z}(t)=\alpha\cos(\omega_M t)$ with the phase modulation
amplitude $\alpha$ being a zero's of $\mathrm{J_0}$.

\begin{figure}[t]
\includegraphics[width=10cm]{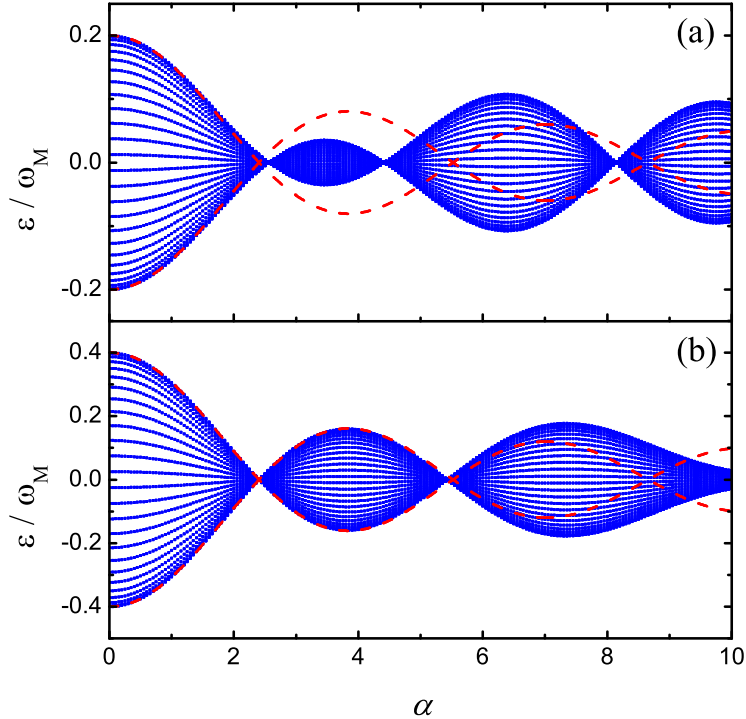}
\caption{(color online) Quasi-energies as a function of
phase-modulation strength $\alpha$, for the Hamiltonian of Eq.
(\ref{hamnonrwack}), with $\phi_x(t) = \alpha \cos(\omega_M t)$.
$V = 0.2 \Omega$. (a) $\Omega = 2 \omega_M$. (b) $\Omega = 4
\omega_M$. Each blue curve corresponds to a different wavevector
$k_x$, in the range of $-\pi/a < k_x < \pi/a$.  Dashed red line is
the envelope for the same Hamiltonian, but calculated using the
rotating wave approximation.}\label{figBCnonrwa}
\end{figure}

We confirm the intuitive analysis above by calculating the Floquet
bandstructure in the case where $V = 0.2\Omega$, and hence the
rotating wave approximation is no longer valid (blue lines in
Figure \ref{figBCnonrwa}), and by comparing such calculations to
the prediction of the range of quasi-energies with rotating wave
approximation. (Red lines in Figure \ref{figBCnonrwa}). Figure
\ref{figBCnonrwa}(a) shows the case with $\Omega = 2 \omega_M$.
Introducing the counter rotating term indeed modifies the
bandstructure. Nevertheless, the bandwidth still collapses near a
phase modulation strength of $\alpha = 2. 40483$. Thus, dynamic
localization still occurs in this system beyond the rotating wave
approximation.  Figure \ref{figBCnonrwa}(b) shows the case with
$\Omega = 4 \omega_M$. Comparing Figures \ref{figBCnonrwa}(a) and
\ref{figBCnonrwa}(b), we observe that the discrepancy in the
bandstructures between the cases with or without rotating wave
approximation becomes smaller as $\omega_M$ is reduced, in spite
of the fact that with $V = 0.2 \Omega$ we are significantly
outside the regime where the rotating wave approximation is valid.
This observation is consistent with the analytic results derived
above based on instantaneous bandstructure.

Experimentally, effective gauge field for photons have already
been experimentally observed using two modulators \cite{tzuang14}.
The demonstration of the theoretical proposal here requires
further integration of larger numbers of modulators. The
experimental feasibility of such integration has been discussed in
Ref. \cite{fang12np}. While for illustration purpose we have
focused on a photonic gauge potential through the use of temporal
refractive index modulation, the concept here should be relevant
for other proposals of photonic gauge potential as well, include
those based on magneto-optical effects
\cite{umucalilar11,fangpra13}, as well as spin-dependent photonic
gauge potential
\cite{hafezi11,hafezi13,khanikaev13,rechstman13,mittal14} and
optomechnanical gauge potential
\cite{schmidtarxiv13,peanoarxiv14}. In summary, we have shown that
three-dimensional dynamic localization of light can be achieved
with an effective gauge potential for photons. The results provide
additional evidence of the exciting prospects of photonic gauge
potential for the control of light propagation.

\begin{acknowledgments}
This work is supported in part by U.S. Air Force Office of
Scientific Research Grant No. FA9550-09-1-0704 and U.S. National
Science Foundation Grant No. ECCS- 1201914.
\end{acknowledgments}

\end{document}